\documentclass[preprint,prd,aps,showpacs,showkeys,nofootinbib]{revtex4}
\usepackage{graphicx}
\usepackage{dcolumn}
\usepackage{bm}
\usepackage{color}
\usepackage[dvipsnames]{xcolor}
\usepackage{amssymb}

\definecolor{light-gray}{gray}{0.78}
\definecolor{mid-gray}{gray}{0.55}
\definecolor{dark-gray}{gray}{0.32}

\begin{document}
\title{Study the Higgs mass with the effective potential and Higgs decays in the $U(1)_X$SSM}
\author{Shu-Min Zhao$^{1,2}$\footnote{zhaosm@hbu.edu.cn},  Xi Wang$^{1,2}$, Xing-Xing Dong$^{1,2}$, Hai-Bin Zhang$^{1,2}$,
Tai-Fu Feng$^{1,2,3}$}

\affiliation{$^1$ Department of Physics, Hebei University, Baoding 071002, China}
\affiliation{$^2$ Key Laboratory of High-precision Computation and Application of Quantum Field Theory of Hebei Province, Baoding 071002, China}
\affiliation{$^3$ Department of Physics, Chongqing University, Chongqing 401331, China}

\date{\today}
\begin{abstract}
 As the U(1) extension of the minimal supersymmetric standard model, the $U(1)_X$SSM has new super fields
 such as right-handed neutrinos and three Higgs singlets. In the $U(1)_X$SSM, the lightest CP-even Higgs mass $m_{h^0}$ is researched
  through the Higgs effective potential with one loop corrections. We also calculate the Higgs decays
  $h^0\rightarrow\gamma\gamma$, $h^0\rightarrow VV~(V=W,~Z)$, $h^0\rightarrow l\bar{l}Z$ and $h^0\rightarrow \nu\bar{\nu}Z$.
  The obtained results are reasonable, which are in favour of the study of the Higgs characteristic and the phenomenology of the $U(1)_X$SSM.
\end{abstract}



\maketitle

\section{Introduction}

In the standard model (SM) of particle physics, the Higgs boson is the last particle
discovered and inherently related to
the mechanism of spontaneous symmetry breaking (SSB). The observation of the Higgs
boson in 2012 \cite{2012Higgs} is a great success  of the SM.
Nevertheless, more detailed and precise investigations of Higgs are required
 and the search for new physics beyond the SM is one of the major issues of particle physics.
 On the other hand, the SM has some shortcomings, such as can not explain neutrino mass and mixing\cite{neutrino1, neutrino2},
 can not provide the candidates for cold dark matter, can not explain the asymmetry of matter and antimatter in the universe, etc..

A famous extension of the SM is the minimal supersymmetric extension of the standard model (MSSM)\cite{MSSM},
which has been researched by physicists for several decades.
People also extend MSSM into multiple models, in which the U(1) extensions of MSSM are interesting.
 The $U(1)_X$ extension of MSSM is called as $U(1)_X$SSM \cite{Sarah,ZSMJHEP20,ZSMJHEP22} with
 the local gauge group $SU(3)_C\times SU(2)_L \times U(1)_Y \times U(1)_X$.
 $U(1)_X$SSM has more superfields (three Higgs singlets and right-handed neutrinos) than MSSM.
  The added right-handed neutrinos can not only explain the tiny mass of neutrino, but also provide a new dark matter candidate-light sneutrino.
    The $\mu$ problem appearing in the MSSM is relieved in the $U(1)_X$SSM by the terms $\mu\hat{H}_u\hat{H}_d$ and $\lambda_H\hat{S}\hat{H}_u\hat{H}_d$  producing an effective $\mu_{eff}=\mu+\lambda_Hv_S/\sqrt{2}$.
 The Higgs singlet $S$ has a non-zero VEV ($v_S/\sqrt{2}$).
 The mixing of the CP-even parts of $H_d,~H_u,~\eta,~\bar{\eta},~S$ can improve the lightest CP-even Higgs mass at tree level.

  Higgs pairs can be produced through gluon-gluon fusion \cite{ggh,RPTHiggs} in
pp collision through loop diagrams.
The Higgs boson mass and decays including $h^0\rightarrow \gamma\gamma$ and $h^0\rightarrow VV~ ( {\rm with}~ V=Z,~W)$
have been studied in several models such as MSSM, NMSSM\cite{cao}, B-LSSM\cite{B-LHiggs}, BLMSSM\cite{TFNPB} and so on.
In the $U(1)_X$SSM, we study the lightest CP-even Higgs mass through the Higgs effective potential with one loop corrections.
The Higgs boson decays  $h^0\rightarrow \gamma\gamma,~h^0\rightarrow VV~( {\rm with}~ V=Z,~W),~h^0\rightarrow l\bar{l} Z,~h^0\rightarrow \nu_l\bar{\nu}_l Z~({\rm with} ~l=e,~\mu,~\tau)$
are all calculated in this work.

 For the decays $h^0$ to $\gamma\gamma,~ ZZ$ and $WW$,
 the current values of the corresponding ratios $R_{\gamma\gamma},~R_{ZZ},~R_{WW}$ are
 respectively  $R_{\gamma\gamma}=1.10\pm0.07,~R_{WW}=1.19\pm0.12$ and $R_{ZZ}=1.01\pm0.07$\cite{2022pdg}.
 $R_{XX}$ is the ratio defined as
\begin{eqnarray}
R_{XX}=\frac{\Gamma_{U(1)_X}(h^0\rightarrow gg)\Gamma_{U(1)_X}(h^0\rightarrow XX)}{\Gamma_{SM}(h^0\rightarrow gg)\Gamma_{SM}(h^0\rightarrow XX)}.
\end{eqnarray}

For the Higgs boson decays $h^0\rightarrow l\bar{l} Z,~h^0\rightarrow \nu_l\bar{\nu}_l Z~({\rm with} ~l=e,~\mu,~\tau)$\cite{QCF}, they are in the
reachable region of LHC.
Some future experiments including two circular lepton colliders(CEPC and FCC-ee)\cite{QQ} and a linear
 lepton collider(ILC) have been proposed to study the properties of the Higgs boson.
 The accuracy of these colliders in measuring the Higgs boson decays will be improved obviously, and we believe
 that the decays $h^0\rightarrow l\bar{l} Z,~h^0\rightarrow \nu_l\bar{\nu}_l Z~( {\rm with}~l=e,~\mu,~\tau)$
can be detected in the near future.

In section 2, we briefly introduce the main content of $U(1)_X$SSM and its superfields. The formulation for the Higgs effective
potential and Higgs boson decays $h^0\rightarrow \gamma\gamma~(WW,~ ZZ)$,
$h^0\rightarrow l\bar{l}Z,~h^0\rightarrow \nu_l\bar{\nu}_lZ~( {\rm with}~l=e,~\mu,~\tau)$ are shown in the section 3.
We analyse the results numerically in the section 4, and obtain reasonable parameter space.
The last section is used for the discussion and conclusion.

\section{The main content of $U(1)_X$SSM}

We extend MSSM with the local gauge group $U(1)_X$ to obtain $U(1)_X$SSM,
which has new super fields: three generation right-handed neutrinos and three Higgs singlets. Then $U(1)_X$SSM can account for
the data of neutrino oscillation. The introduction of three Higgs singlets ($\eta,~\bar{\eta}$ and $S$)
leads to the extension of mass squared matrix for CP-even Higgs.
The new mixing of Higgs can  improve the lightest CP-even Higgs mass at the tree level.
One can find the particle contents in the work\cite{ZSMJHEP20,WTTJHEP}.

In $U(1)_X$SSM, the superpotential and soft SUSY breaking terms are shown here\cite{ZSMJHEP20,ZSMJHEP22,WTTJHEP}
\begin{eqnarray}
&&W=l_W\hat{S}+\mu\hat{H}_u\hat{H}_d+M_S\hat{S}\hat{S}-Y_d\hat{d}\hat{q}\hat{H}_d-Y_e\hat{e}\hat{l}\hat{H}_d+\lambda_H\hat{S}\hat{H}_u\hat{H}_d
\nonumber\\&&+\lambda_C\hat{S}\hat{\eta}\hat{\bar{\eta}}+\frac{\kappa}{3}\hat{S}\hat{S}\hat{S}+Y_u\hat{u}\hat{q}\hat{H}_u+Y_X\hat{\nu}\hat{\bar{\eta}}\hat{\nu}
+Y_\nu\hat{\nu}\hat{l}\hat{H}_u.
\\
&&\mathcal{L}_{soft}=\mathcal{L}_{soft}^{MSSM}-B_SS^2-L_SS-\frac{T_\kappa}{3}S^3-T_{\lambda_C}S\eta\bar{\eta}
+\epsilon_{ij}T_{\lambda_H}SH_d^iH_u^j\nonumber\\&&
-T_X^{IJ}\bar{\eta}\tilde{\nu}_R^{*I}\tilde{\nu}_R^{*J}
+\epsilon_{ij}T^{IJ}_{\nu}H_u^i\tilde{\nu}_R^{I*}\tilde{l}_j^J
-m_{\eta}^2|\eta|^2-m_{\bar{\eta}}^2|\bar{\eta}|^2\nonumber\\&&
-m_S^2S^2-(m_{\tilde{\nu}_R}^2)^{IJ}\tilde{\nu}_R^{I*}\tilde{\nu}_R^{J}
-\frac{1}{2}\Big(M_S\lambda^2_{\tilde{X}}+2M_{BB^\prime}\lambda_{\tilde{B}}\lambda_{\tilde{X}}\Big)+h.c~~.
\end{eqnarray}

The two Higgs doublets and three Higgs singlets are
\begin{eqnarray}
&&H_{u}=\left(\begin{array}{c}H_{u}^+\\{1\over\sqrt{2}}\Big(v_{u}+\phi_{u}+iP_{u}^0\Big)\end{array}\right),
~~~~~~
H_{d}=\left(\begin{array}{c}{1\over\sqrt{2}}\Big(v_{d}+\phi_{d}+iP_{d}^0\Big)\\H_{d}^-\end{array}\right),
\nonumber\\
&&\eta={1\over\sqrt{2}}\Big(v_{\eta}+\phi_{\eta}+iP_{\eta}^0\Big),~~~~~~~~~~~~~~~
\bar{\eta}={1\over\sqrt{2}}\Big(v_{\bar{\eta}}+\phi_{\bar{\eta}}+iP_{\bar{\eta}}^0\Big),\nonumber\\&&
\hspace{4.0cm}S={1\over\sqrt{2}}\Big(v_{S}+\phi_{s}+iP_{s}^0\Big).
\end{eqnarray}
$v_u,~v_d,~v_\eta$,~ $v_{\bar\eta}$ and $v_S$ respectively represent the VEVs of the Higgs super fields $H_u$, $H_d$, $\eta$, $\bar{\eta}$ and $S$. The definitions of two angles are $\tan\beta=v_u/v_d$ and $\tan\beta_\eta=v_{\bar{\eta}}/v_{\eta}$.

 $Y^Y$ denotes the $U(1)_Y$ charge and $Y^X$ represents the $U(1)_X$ charge.
One can write the covariant derivatives of $U(1)_X$SSM in the form
\begin{eqnarray}
&&D_\mu=\partial_\mu-i\left(\begin{array}{cc}Y,&X\end{array}\right)
\left(\begin{array}{cc}g_{Y},&g{'}_{{YX}}\\g{'}_{{XY}},&g{'}_{{X}}\end{array}\right)
\left(\begin{array}{c}A_{\mu}^{\prime Y} \\ A_{\mu}^{\prime X}\end{array}\right)\;,
\label{gauge1}
\end{eqnarray}
 where $A_{\mu}^{\prime Y}$ and $A^{\prime X}_\mu$ denote the gauge fields of $U(1)_Y$ and $U(1)_X$ respectively.

It is convenient to perform a change of the basis with the rotation
matrix $R$\cite{Rmatrix}
\begin{eqnarray}
&&D_\mu=\partial_\mu-i\left(\begin{array}{cc}Y^Y,&Y^X\end{array}\right)
\left(\begin{array}{cc}g_{Y},&g{'}_{{YX}}\\g{'}_{{XY}},&g{'}_{{X}}\end{array}\right)R^TR
\left(\begin{array}{c}A_{\mu}^{\prime Y} \\ A_{\mu}^{\prime X}\end{array}\right)\;,
\\
&&\left(\begin{array}{cc}g_{Y},&g{'}_{{YX}}\\g{'}_{{XY}},&g{'}_{{X}}\end{array}\right)
R^T=\left(\begin{array}{cc}g_{1},&g_{{YX}}\\0,&g_{{X}}\end{array}\right),~~~
R\left(\begin{array}{c}A_{\mu}^{\prime Y} \\ A_{\mu}^{\prime X}\end{array}\right)
=\left(\begin{array}{c}A_{\mu}^{Y} \\ A_{\mu}^{X}\end{array}\right)\;.
\end{eqnarray}
In the end, the covariant derivatives of the $U(1)_X$SSM turn into
\begin{eqnarray}
&&D_\mu=\partial_\mu-i\left(\begin{array}{cc}Y^Y,&Y^X\end{array}\right)
\left(\begin{array}{cc}g_{1},&g_{{YX}}\\0,&g_{{X}}\end{array}\right)
\left(\begin{array}{c}A_{\mu}^{Y} \\ A_{\mu}^{X}\end{array}\right)\;.
\end{eqnarray}

In the $U(1)_X$SSM, the gauge bosons $A^{X}_\mu,~A^Y_\mu$ and $V^3_\mu$ mix together at the tree level.
We deduce their mass eigenvalues as
\begin{eqnarray}
&&m_\gamma^2=0,\nonumber\\
&&m_{Z,{Z^{'}}}^2=\frac{1}{8}\Big((g_{1}^2+g_2^2+(g_{YX}+g_X)^2)v^2+4g_{X}^2\xi^2 \nonumber\\
&&\mp\sqrt{(g_{1}^2+g_{2}^2+(g_{YX}+g_X)^2)^2v^4+8((g_{YX}+g_X)^2-g_{1}^2-
g_{2}^2)g_{X}^2v^2\xi^2+16g_{X}^4\xi^4}\Big).
\end{eqnarray}
with $v^2=v_u^2+v_d^2$ and $\xi^2=v_\eta^2+v_{\bar{\eta}}^2$. Two mixing
angles $\theta_{W}$ and $\theta_{W}'$\cite{ZSMJHEP22,WTTJHEP} are used here.

Supposing $\mu,~\lambda_H,~\lambda_C,~l_W,~ M_S,~B_\mu,~L_S,~T_{\kappa},~ T_{\lambda_C},~ T_{\lambda_H},~\kappa,~B_S$  as real parameters, we show the simplified Higgs potential at tree level\cite{ZSMJHEP20}
\begin{eqnarray}
&&V_{0}=\frac{1}{2}g_X(g_X+g_{YX})(|H_d^0|^2-|H_u^0|^2)(|\eta|^2-|\bar{\eta}|^2)+\lambda_H^2|H_u^0H_d^0|^2+m^2_{S}|S|^2+l_W^2\nonumber\\&&
+\frac{1}{8}\Big(g_1^2+g_2^2+(g_X+g_{YX})^2\Big)(|H_d^0|^2-|H_u^0|^2)^2+\frac{1}{2}g_X^2(|\eta|^2-|\bar{\eta}|^2)^2+\lambda_C^2|\eta\bar{\eta}|^2
\nonumber\\&&
+(\mu^2+\lambda_H^2|S|^2+2\mathrm{Re}[\mu\lambda_HS])(|H_d^0|^2+|H_u^0|^2)+\lambda_C^2|S|^2(|\eta|^2+|\bar{\eta}|^2)+m^2_\eta|\eta|^2
\nonumber\\&&+2\mathrm{Re}[(l_W+2 M_SS^*)(\lambda_C\eta\bar{\eta}-\lambda_HH_u^0H_d^0
+\kappa S^2)]+4M_S^2|S|^2+\kappa^2|S|^4+m^2_{\bar{\eta}}|\bar{\eta}|^2\nonumber\\&&+
2\mathrm{Re}[\lambda_C\kappa\eta^*\bar{\eta}^*S^2+2l_WM_SS
-\lambda_C\lambda_H\eta^*\bar{\eta}^*H_u^0H_d^0]+
m^2_{H_u^0}|H_u|^2+m^2_{H_d}|H_d|^2\nonumber\\&&
+ 2\mathrm{Re}\Big[L_S S- H_d^0H_u^0\Big(B_\mu+\lambda_H\kappa  (S^2)^*+T_{\lambda_H} S\Big)
+\frac{1}{3}T_kS^3+ T_{\lambda_C}\eta\bar{\eta}S+B_S S^2\Big].\label{Vtree}
\end{eqnarray}
The corresponding tadpole equations at tree level are also obtained in Ref.\cite{ZSMJHEP20}.
The tree level mass squared matrix for CP-even Higgs $({\phi}_{d}, {\phi}_{u}, {\phi}_{\eta}, {\phi}_{\overline{\eta}}, {\phi}_{s})$ is
\begin{eqnarray}
M^2_{h,tree} = \left(
\begin{array}{ccccc}
m_{{\phi}_{d}{\phi}_{d}} &m_{{\phi}_{u}{\phi}_{d}} &m_{{\phi}_{\eta}{\phi}_{d}} &m_{{\phi}_{\bar{\eta}}{\phi}_{d}} &m_{{\phi}_{s}{\phi}_{d}}\\
m_{{\phi}_{d}{\phi}_{u}} &m_{{\phi}_{u}{\phi}_{u}} &m_{{\phi}_{\eta}{\phi}_{u}} &m_{{\phi}_{\bar{\eta}}{\phi}_{u}} &m_{{\phi}_{s}{\phi}_{u}}\\
m_{{\phi}_{d}{\phi}_{\eta}} &m_{{\phi}_{u}{\phi}_{\eta}} &m_{{\phi}_{\eta}{\phi}_{\eta}} &m_{{\phi}_{\bar{\eta}}{\phi}_{\eta}} &m_{{\phi}_{s}{\phi}_{\eta}}\\
m_{{\phi}_{d}{\phi}_{\bar{\eta}}} &m_{{\phi}_{u}{\phi}_{\bar{\eta}}} &m_{{\phi}_{\eta}{\phi}_{\bar{\eta}}} &m_{{\phi}_{\bar{\eta}}{\phi}_{\bar{\eta}}} &m_{{\phi}_{s}{\phi}_{\bar{\eta}}}\\
m_{{\phi}_{d}{\phi}_{s}} &m_{{\phi}_{u}{\phi}_{s}} &m_{{\phi}_{\eta}{\phi}_{s}} &m_{{\phi}_{\bar{\eta}}{\phi}_{s}} &m_{{\phi}_{s}{\phi}_{s}}\end{array}
\right),\label{Rsneu}
 \end{eqnarray}
\begin{eqnarray}
&&m_{\phi_{d}\phi_{d}}= m_{H_d}^2+  \mu^2
 +\frac{1}{8} \Big( [g_{1}^{2}+(g_{X}+g_{YX})^{2}+g_2^2] (3 v_{d}^{2}  - v_{u}^{2})\nonumber \\
&&\hspace{1.5cm}+2 (g_{Y X} g_{X}+g_X^2) ( v_{\eta}^{2}- v_{\bar{\eta}}^{2})\Big)+ \sqrt{2} v_S \mu {\lambda}_{H}+\frac{1}{2} (v_{u}^{2} + v_S^{2}){\lambda}_{H}^2,
\\&&m_{\phi_{d}\phi_{u}} = -\frac{1}{4} \Big(g_{2}^{2} + (g_{Y X} + g_{X})^{2} + g_1^{2}\Big)v_d v_u
 + {\lambda}_{H}^2 v_d v_u - {\lambda}_{H} l_W \nonumber \\
&&\hspace{1.5cm}-\frac{1}{2}{\lambda}_{H} (v_{\eta} v_{\bar{\eta}} {\lambda}_{C}  + v_S^{2} \kappa )
 - B_{\mu}- \sqrt{2} v_S (\frac{1}{2}T_{{\lambda}_{H}}  + M_S {\lambda}_{H} ),
\\ &&m_{\phi_{u}\phi_{u}} = m_{H_u}^2+ \mu^2+\frac{1}{8} \Big( [g_{1}^{2}+(g_{X}+g_{YX})^{2}+g_2^2] (3 v_{u}^{2}  - v_{d}^{2})\nonumber \\
&&\hspace{1.5cm}+2 (g_{Y X} g_{X}+g_X^2) ( v_{\bar{\eta}}^{2}-v_{\eta}^{2})\Big)
 +  \sqrt{2} v_S\mu {\lambda}_{H}   + \frac{1}{2}(v_{d}^{2} + v_S^{2}){\lambda}_{H}^2,
\\&&m_{\phi_{d}\phi_{\eta}} = \frac{1}{2}g_{X} (g_{Y X} + g_{X})v_d v_{\eta}
  -\frac{1}{2} v_u v_{\bar{\eta}} {\lambda}_{H} {\lambda}_{C} ,
\\&&m_{\phi_{u}\phi_{\eta}} = -\frac{1}{2}g_{X} (g_{Y X} + g_{X})v_u v_{\eta}
-\frac{1}{2} v_d v_{\bar{\eta}} {\lambda}_{H} {\lambda}_{C},
\\&&m_{\phi_{\eta}\phi_{\eta}} = m_{\eta}^2 +\frac{1}{4} \Big((g_{Y X} g_{X}+g_X^2) ( v_{d}^{2}
- v_{u}^{2})+2g_{X}^{2}
( 3 v_{\eta}^{2}-v_{\bar{\eta}}^{2})\Big)+\frac{{\lambda}_{C}^2}{2} (v_{\bar{\eta}}^{2} + v_S^{2}),
\\&&m_{\phi_{d}\phi_{\bar{\eta}}} = -\frac{1}{2}g_{X} (g_{Y X} + g_{X})v_d v_{\bar{\eta}}
  -\frac{1}{2} v_u v_{\eta} {\lambda}_{H} {\lambda}_{C},
\\&&m_{\phi_{u}\phi_{\bar{\eta}}} = \frac{1}{2}g_{X} (g_{Y X} + g_{X})v_u v_{\bar{\eta}}
 -\frac{1}{2} v_d v_{\eta} {\lambda}_{H} {\lambda}_{C},
\\&&m_{\phi_{\eta}\phi_{\bar{\eta}}} = ({\lambda}_{C}^2- g_{X}^{2})v_{\eta} v_{\bar{\eta}}+\frac{{\lambda}_{C}}{2}(2 l_W  - {\lambda}_{H} v_d v_u )
+ \frac{v_S }{\sqrt{2}} (2 M_S {\lambda}_{C} + T_{{\lambda}_{C}}) + \frac{v_S^{2}}{2} {\lambda}_{C} \kappa,
\\&&m_{\phi_{\bar{\eta}}\phi_{\bar{\eta}}} = m_{\bar{\eta}}^2+\frac{1}{4} \Big((g_{Y X} g_{X}+g_X^2)
 ( v_{u}^{2}- v_{d}^{2})+2g_{X}^{2}( 3 v_{\bar{\eta}}^{2}-v_{\eta}^{2})\Big)+\frac{{\lambda}_{C}^2 }{2} \Big(v_{\eta}^{2} + v_S^{2}\Big),
\\&&m_{\phi_{d}{\phi}_{s}} = \Big({\lambda}_{H} v_d v_S  + \sqrt{2} v_d \mu  -  v_u ( \kappa v_S
 + \sqrt{2} M_S )\Big){\lambda}_{H} - \frac{1}{\sqrt{2}}v_u T_{{\lambda}_{H}},
\\&&m_{\phi_{u}{\phi}_{s}} =  \Big( {\lambda}_{H} v_u v_S  + \sqrt{2} v_u \mu
-v_d (\kappa v_S  + \sqrt{2} M_S )\Big){\lambda}_{H}
- \frac{1}{\sqrt{2}}  v_dT_{{\lambda}_{H}},
\\&&m_{\phi_{\eta}{\phi}_{s}} = \Big( {\lambda}_{C} v_{\eta} v_S  + v_{\bar{\eta}} (\kappa v_S
 + \sqrt{2} M_S )\Big){\lambda}_{C}  +\frac{1}{\sqrt{2}}v_{\bar{\eta}} T_{{\lambda}_{C}},
\\&&m_{\phi_{\bar{\eta}}{\phi}_{s}} =\Big( {\lambda}_{C} v_{\bar{\eta}} v_S  + v_{\eta}(\kappa v_S
 + \sqrt{2} M_S )\Big){\lambda}_{C} + \frac{1}{\sqrt{2}}v_{\eta}
 T_{{\lambda}_{C}},
\\&&m_{{\phi}_{s}{\phi}_{s}} = m^2_{S}+ \Big(2 l_W  + 3v_S (\kappa v_S  + 2\sqrt{2} M_S )
+ {\lambda}_{C} v_{\eta} v_{\bar{\eta}}  - {\lambda}_{H} v_d v_u \Big)\kappa+2 {B_{S}}\nonumber \\
&&\hspace{1.5cm} +\frac{1}{2}{\lambda}_{C}^2 \xi^2+\frac{1}{2}{\lambda}_{H}^2 v^{2}
  + 4 M_S^2   + \sqrt{2} v_S T_{\kappa}.
\end{eqnarray}

\section{formulation}

The one loop effective potential  can be written in the following form
\begin{eqnarray}
V_{eff}=V_{0}+V_{1}.
\end{eqnarray}
Here, $V_{1}$ is the potential from one loop  correction. With the dimensional reduction and the DR renormalization scheme, the
effective Higgs potential up to one loop correction is shown in Landau gauge, and
the concrete form of $V_{1}$ is\cite{oneloopT0V, LiTJ,YBJPG}
\begin{eqnarray}
&&V_{1}=\sum_i\frac{n_i}{64\pi^2}m_i^4(\phi_d,\phi_u,\phi_\eta, \phi_{\bar{\eta}},\phi_s)\Big(
\log\frac{m_i^2(\phi_d,\phi_u,\phi_\eta, \phi_{\bar{\eta}},\phi_s)}{Q^2}-\frac{3}{2}\Big).
\end{eqnarray}
We take the renormalization
scale Q  at TeV order. The degrees of freedom for each mass eigenstate are represented by $n_i$
(-12 for quarks, -4 for leptons and charginos, -2 for neutralinos and neutrinos,
6 for squarks, 2 for sleptons and charged Higgs, 3 and 6 for $Z(Z^\prime)$ and $W$ bosons,
 1 for sneutrinos and the neutral Higgs scalars). The mass matrices are needed,
  and we collect the mass matrices of CP-even sneutrino, CP-odd sneutrino, slepton, squark, chargino and neutralino.

The mass matrix for CP-even sneutrino $({\phi}_{l}, {\phi}_{r})$ reads
\begin{eqnarray}
M^2_{\tilde{\nu}^R} = \left(
\begin{array}{cc}
m_{{\phi}_{l}{\phi}_{l}} &m^T_{{\phi}_{r}{\phi}_{l}}\\
m_{{\phi}_{l}{\phi}_{r}} &m_{{\phi}_{r}{\phi}_{r}}\end{array}
\right),\label{Rsneu}
 \end{eqnarray}
\begin{eqnarray}
&&m_{{\phi}_{l}{\phi}_{l}}= \frac{1}{8} \Big((g_{1}^{2} + g_{Y X}^{2} + g_{2}^{2}+ g_{Y X} g_{X})( v_{d}^{2}- v_{u}^{2})
+  2g_{Y X} g_{X}(v_{\eta}^{2}- v_{\bar{\eta}}^{2})\Big)
+\frac{ v_{u}^{2}}{2}{Y_{\nu}^2}  + m_{\tilde{L}}^2\nonumber,
 \\&&m_{{\phi}_{l}{\phi}_{r}} = \frac{1}{\sqrt{2} } v_uT_\nu  +  v_u v_{\bar{\eta}} {Y_X  Y_\nu}
  - \frac{1}{2}v_d ({\lambda}_{H}v_S  + \sqrt{2} \mu )Y_\nu,\nonumber\\&&
m_{{\phi}_{r}{\phi}_{r}}= \frac{1}{8} \Big((g_{Y X} g_{X}+g_{X}^{2})(v_{d}^{2}- v_{u}^{2})
+2g_{X}^{2}(v_{\eta}^{2}- v_{\bar{\eta}}^{2})\Big) + v_{\eta} v_S Y_X {\lambda}_{C}\nonumber \\&&\hspace{1.8cm}
 +m_{\tilde{\nu}}^2 + \frac{1}{2} v_{u}^{2}|Y_\nu|^2+  v_{\bar{\eta}} (2 v_{\bar{\eta}}Y_X ^2  + \sqrt{2} T_X).
\end{eqnarray}
To obtain the masses of sneutrinos, the rotation matrix $Z^R$ is used to diagonalize $M^2_{\tilde{\nu}^R}$.

We also deduce the mass matrix for CP-odd sneutrino $({\sigma}_{l}, {\sigma}_{r})$
\begin{eqnarray}
M^2_{\tilde{\nu}^I} = \left(
\begin{array}{cc}
m_{{\sigma}_{l}{\sigma}_{l}} &m^T_{{\sigma}_{r}{\sigma}_{l}}\\
m_{{\sigma}_{l}{\sigma}_{r}} &m_{{\sigma}_{r}{\sigma}_{r}}\end{array}
\right),
 \end{eqnarray}
\begin{eqnarray}
&&m_{{\sigma}_{l}{\sigma}_{l}}= \frac{1}{8} \Big((g_{1}^{2} + g_{Y X}^{2} + g_{2}^{2}+  g_{Y X} g_{X})( v_{d}^{2}- v_{u}^{2})
+  2g_{Y X} g_{X}(v_{\eta}^{2}-v_{\bar{\eta}}^{2})\Big)
+\frac{ v_{u}^{2}}{2}Y_{\nu}^2  + m_{\tilde{L}}^2\nonumber,
 \\&&m_{{\sigma}_{l}{\sigma}_{r}} = \frac{1}{\sqrt{2} } v_uT_\nu -  v_u v_{\bar{\eta}} {Y_X  Y_\nu}
  - \frac{1}{2}v_d ({\lambda}_{H}v_S  + \sqrt{2} \mu )Y_\nu,\nonumber\\&&
m_{{\sigma}_{r}{\sigma}_{r}}= \frac{1}{8} \Big((g_{Y X} g_{X}+g_{X}^{2})(v_{d}^{2}- v_{u}^{2})
+2g_{X}^{2}(v_{\eta}^{2}- v_{\bar{\eta}}^{2})\Big)- v_{\eta} v_S Y_X {\lambda}_{C}\nonumber \\&&\hspace{1.8cm}
+m_{\tilde{\nu}}^2 + \frac{1}{2} v_{u}^{2}|Y_\nu|^2+  v_{\bar{\eta}} (2 v_{\bar{\eta}}Y_X  Y_X  - \sqrt{2} T_X).
\end{eqnarray}
 We use $Z^I$ to diagonalize the mass squared matrix of the sneutrino $M^2_{\tilde{\nu}^I}$.

In the basis $(\tilde{e}_L, \tilde{e}_R)$, the mass matrix for slepton is shown and diagonalized by $Z^E$ through the
formula $Z^E m^2_{\tilde{e}} Z^{E,\dagger} = m^{diag}_{2,\tilde{e}}$,
\begin{equation}
m^2_{\tilde{e}} = \left(
\begin{array}{cc}
m_{\tilde{e}_L\tilde{e}_L^*} &\frac{1}{2} \Big(\sqrt{2} v_d T_{e}^{\dagger}  - v_u ({\lambda}_{H} v_S  + \sqrt{2} \mu )Y_{e}^{\dagger} \Big)\\
\frac{1}{2} \Big(\sqrt{2} v_d T_e  - v_u Y_e (\sqrt{2} \mu  + v_S {\lambda}_{H} )\Big) &m_{\tilde{e}_R\tilde{e}_R^*}\end{array}
\right).
 \end{equation}
\begin{eqnarray}
&&m_{\tilde{e}_L\tilde{e}_L^*} = m_{\tilde{L}}^2+\frac{1}{8} \Big((g_{1}^{2} + g_{Y X}^{2}
+ g_{Y X} g_{X} -g_2^2)(v_{d}^{2}- v_{u}^{2})+ 2 g_{Y X} g_{X}( v_{\eta}^{2}- v_{\bar{\eta}}^{2}
)
\Big)+\frac{v_{d}^{2}}{2} {Y_{e}^2} ,\nonumber\\&&
m_{\tilde{e}_R\tilde{e}_R^*} = m_{\tilde{E}}^2-\frac{1}{8}  \Big([2(g_{1}^{2} + g_{Y X}^{2})+3g_{Y X} g_{X}+g_{X}^{2}]
( v_{d}^{2}- v_{u}^{2})\nonumber\\&&\hspace{1.7cm}+(4g_{Y X} g_{X}+2g_{X}^{2})(v_{\eta}^{2}- v_{\bar{\eta}}^{2})
\Big)+\frac{1}{2} v_{d}^{2} {  Y_{e}^2}.
\end{eqnarray}

The  squared mass matrix for down type squark is shown
 in the basis $\left(\tilde{d}^0_{L}, \tilde{d}^0_{R}\right)$
\begin{equation}
M^2_{\tilde{D}} = \left(
\begin{array}{cc}
m_{\tilde{d}_L^0\tilde{d}_L^{0,*}} &m^\dagger_{\tilde{d}_R^0\tilde{d}_L^{0,*}}\\
m_{\tilde{d}_L^0\tilde{d}_R^{0,*}} &m_{\tilde{d}_R^0\tilde{d}_R^{0,*}}\end{array}
\right),
\end{equation}
where
\begin{eqnarray}
&&m_{\tilde{d}_L^0\tilde{d}_L^{0,*}} = \frac{1}{24}
 \Big( (3 g_{2}^{2}  + g_{1}^{2} + g_{Y X}^{2}+  g_{Y X} g_{X}) ( v_{u}^{2}- v_{d}^{2}  ) +  2g_{Y X} g_{X}
 ( v_{\bar{\eta}}^{2}  - v_{\eta}^{2})\Big)+m_{\tilde{Q}}^2  +\frac{ v_{d}^{2}}{2} {Y_{d}^2},\nonumber\\
&&m_{\tilde{d}_L^0\tilde{d}_R^{0,*}} = -\frac{1}{2}  \Big(\sqrt{2}  (- v_d T_d  + v_u Y_d \mu ) + v_u v_S Y_d {\lambda}_{H} \Big),\nonumber\\
&&m_{\tilde{d}_R^0\tilde{d}_R^{0,*}} = \frac{1}{24}   \Big(  (2 g_{1}^{2} + 2g_{Y X}^{2}+ 5g_{Y X} g_{X}+3g_{X}^{2})
 ( v_{u}^{2}  - v_{d}^{2})+ 2(2g_{Y X} g_{X}+3 g_{X}^{2} )( v_{\bar{\eta}}^{2}  - v_{\eta}^{2})\Big)\nonumber\\&&\hspace{1.8cm} +m_{\tilde{D}}^2  +\frac{ v_{d}^{2}}{2} {Y_{d}^2}.
\end{eqnarray}

In the basis $\left(\tilde{u}^0_{L}, \tilde{u}^0_{R}\right)$, the squared mass matrix for up type squark is
\begin{equation}
M^2_{\tilde{U}} = \left(
\begin{array}{cc}
m_{\tilde{u}_L^0\tilde{u}_L^{0,*}} &m^\dagger_{\tilde{u}_R^0\tilde{u}_L^{0,*}}\\
m_{\tilde{u}_L^0\tilde{u}_R^{0,*}} &m_{\tilde{u}_R^0\tilde{u}_R^{0,*}}\end{array}
\right),
\end{equation}
where
\begin{eqnarray}
	&&m_{\tilde{u}_L^0\tilde{u}_L^{0,*}} = \frac{1}{24}  \Big( (g_{1}^{2} -3 g_{2}^{2}+ g_{Y X}^{2}+   g_{Y X} g_{X}) ( v_{u}^{2}- v_{d}^{2}  )
 +   g_{Y X} g_{X}  (2 v_{\bar{\eta}}^{2}  -2 v_{\eta}^{2} )\Big)
+  m_{\tilde{Q}}^2  +\frac{ v_{u}^{2}}{2} {Y_{u}^2},\nonumber\\
	&&m_{\tilde{u}_L^0\tilde{u}_R^{0,*}} = -\frac{1}{2}   \Big(\sqrt{2}  (v_d Y_u \mu  - v_u T_u ) + v_d v_S Y_u {\lambda}_{H} \Big),\nonumber\\
	&&m_{\tilde{u}_R^0\tilde{u}_R^{0,*}} = \frac{1}{24}   \Big( (4g_{1}^{2} + 4g_{Y X}^{2}+  7g_{Y X} g_{X}+3  g_{X}^{2})
( v_{d}^{2}- v_{u}^{2})+  2(4g_{Y X} g_{X} +3  g_{X}^{2}) (  v_{\eta}^{2}-v_{\bar{\eta}}^{2}  )\Big) \nonumber \\
	&&\hspace{1.8cm}+  m_{\tilde{U}}^2  +\frac{ v_{u}^{2}}{2} {Y_{u}^2}.\nonumber
\end{eqnarray}

In the basis $(\lambda_{\tilde{B}}, \tilde{W}^0, \tilde{H}_d^0, \tilde{H}_u^0,
\lambda_{\tilde{X}}, \tilde{\eta}, \tilde{\bar{\eta}}, \tilde{s}) $,
the mass matrix for neutralino is,
\begin{equation}
m_{\tilde{\chi}^0} = \left(
\begin{array}{cccccccc}
M_1 &0 &-\frac{g_1}{2}v_d &\frac{g_1}{2}v_u &{M}_{B B'} &0  &0  &0\\
0 &M_2 &\frac{1}{2} g_2 v_d  &-\frac{1}{2} g_2 v_u  &0 &0 &0 &0\\
-\frac{g_1}{2}v_d &\frac{1}{2} g_2 v_d  &0
&m_{\tilde{H}_u^0\tilde{H}_d^0} &m_{\lambda_{\tilde{X}}\tilde{H}_d^0} &0 &0 & - \frac{{\lambda}_{H} v_u}{\sqrt{2}}\\
\frac{g_1}{2}v_u &-\frac{1}{2} g_2 v_u  &m_{\tilde{H}_d^0\tilde{H}_u^0} &0 &m_{\lambda_{\tilde{X}}\tilde{H}_u^0} &0 &0 &- \frac{{\lambda}_{H} v_d}{\sqrt{2}}\\
{M}_{B B'} &0 &m_{\tilde{H}_d^0\lambda_{\tilde{X}}} &m_{\tilde{H}_u^0\lambda_{\tilde{X}}} &{M}_{BL} &- g_{X} v_{\eta}  &g_{X} v_{\bar{\eta}}  &0\\
0  &0 &0 &0 &- g_{X} v_{\eta}  &0 &\frac{1}{\sqrt{2}} {\lambda}_{C} v_S  &\frac{1}{\sqrt{2}} {\lambda}_{C} v_{\bar{\eta}} \\
0  &0 &0 &0 &g_{X} v_{\bar{\eta}}  &\frac{1}{\sqrt{2}} {\lambda}_{C} v_S  &0 &\frac{1}{\sqrt{2}} {\lambda}_{C} v_{\eta} \\
0 &0 & - \frac{{\lambda}_{H} v_u}{\sqrt{2}} &- \frac{{\lambda}_{H} v_d}{\sqrt{2}} &0 &\frac{1}{\sqrt{2}} {\lambda}_{C} v_{\bar{\eta}}
 &\frac{1}{\sqrt{2}} {\lambda}_{C} v_{\eta}  &m_{\tilde{s}\tilde{s}}\end{array}
\right),\label{neutralino}
 \end{equation}

\begin{eqnarray}
&& m_{\tilde{H}_d^0\tilde{H}_u^0} = - \frac{1}{\sqrt{2}} {\lambda}_{H} v_S  - \mu ,~~~~~~~
m_{\tilde{H}_d^0\lambda_{\tilde{X}}} = -\frac{1}{2} (g_{Y X} + g_{X})v_d, \nonumber\\&&
m_{\tilde{H}_u^0\lambda_{\tilde{X}}} = \frac{1}{2} (g_{Y X} + g_{X})v_u
 ,~~~~~~~~~~~~
m_{\tilde{s}\tilde{s}} = 2 M_S  + \sqrt{2} \kappa v_S.\label{neutralino1}
\end{eqnarray}
This matrix is diagonalized by $Z^N$
\begin{equation}
Z^{N*} m_{\tilde{\chi}^0} Z^{N{\dagger}} = m^{diag}_{\tilde{\chi}^0}.
\end{equation}

In the basis $ \left(\tilde{W}^-, \tilde{H}_d^-\right), \left(\tilde{W}^+, \tilde{H}_u^+\right)$, the definition of the mass matrix for charginos is given by
\begin{equation}
M_{\tilde{\chi}^\pm} = \left(
\begin{array}{cc}
M_2 &\frac{1}{\sqrt{2}} g_2 v_u \\
\frac{1}{\sqrt{2}} g_2 v_d  &\frac{1}{\sqrt{2}} {\lambda}_{H} v_S  + \mu\end{array}
\right).
\label{mxzf}
\end{equation}
This matrix is diagonalized by $U$ and $V$
\begin{eqnarray}
U^*M_{\tilde{\chi}^\pm}V^\dagger=M_{\tilde{\chi}^\pm}^{diag}.
\end{eqnarray}

Here, we use the conditions at one loop level through the following formula
\begin{eqnarray}
&&\left \langle \frac{\partial V_{eff}}{\partial \phi_u} \right \rangle=
\left \langle \frac{\partial V_{eff}}{\partial \phi_d} \right \rangle=
\left \langle \frac{\partial V_{eff}}{\partial \phi_\eta} \right \rangle=
\left \langle \frac{\partial V_{eff}}{\partial \phi_{\bar{\eta}}} \right \rangle=
\left \langle \frac{\partial V_{eff}}{\partial \phi_s} \right \rangle=0.
\end{eqnarray}
The corresponding analytic results are very tedious and we resolve the equations numerically.
In order to save space in the text, we do not show the tedious analytic results here.

 The mass squared matrix of CP-even Higgs is corrected by one loop contributions from the effective potential $V_{eff}$
\begin{eqnarray}
M^2_h=M^2_{h,tree}+\Delta M^2_h.
\end{eqnarray}
The elements of the corrected mass squared matrix $M^2_{h,ij}$
can be deduced from the one loop effective potential $V_{eff}$ through the following formula
 \begin{eqnarray}
 M^2_{h,ij}=\Big\langle\frac{\partial^2V_{eff}}{\partial \phi_i \partial \phi_j}\Big|_{\phi_i,\phi_j=\phi_d,\phi_u,\phi_\eta, \phi_{\bar{\eta}},\phi_s}\Big\rangle.
 \end{eqnarray}
The lightest eigenvalue of $M^2_{h}$ should be the square of $m_{h^0}\simeq125$ GeV.

The gluon fusion $(gg\rightarrow h^0)$ \cite{ggh,RPTHiggs} chiefly produces $h^0$ at the LHC.
With large Yukawa coupling, the virtual t quark loop is the dominant contribution during the one loop diagrams.
The large couplings of new particles can lead to considerable corrections
\begin{eqnarray}
&&\Gamma_{{NP}}(h^0\rightarrow gg)={G_{F}\alpha_s^2m_{h^0}^3\over64\sqrt{2}\pi^3}
\Big|\sum\limits_qg_{{h^0 qq}}A_{1/2}(x_q)
+\sum\limits_{\tilde q}g_{h^0\tilde{q}\tilde{q}}{m_{Z}^2\over m_{\tilde q}^2}A_{0}(x_{\tilde{q}})\Big|^2\;,
\label{hgg}
\end{eqnarray}
with $x_a=m_{h^0}^2/(4m_a^2)$. Here, $q$ represents quark, and $\tilde{q}$ denotes squark.
The functions $A_{1/2}(x)$ and $A_0(x)$ are defined as
\begin{eqnarray}
&&A_{1/2}(x)=2\Big(x+(x-1)g(x)\Big)/x^2,~~~~~
A_0(x)=\Big(g(x)-x\Big)/x^2,\\
&&g(x)=\left\{\begin{array}{l}\arcsin^2\sqrt{x},\;x\le1\\
-{1\over4}\Big[\ln{1+\sqrt{1-1/x}\over1-\sqrt{1-1/x}}-i\pi\Big]^2,~~x>1\;.\end{array}\right.
\label{g-function}
\end{eqnarray}

The concrete expressions for $g_{{h^0qq}}$ and $g_{{h^0\tilde{q}\tilde{q}}}$ are
\begin{eqnarray}
&&g_{h^0qq}=\frac{-v}{m_q}C_{h^0qq}~,~~~~~
g_{h^0\tilde{q}\tilde{q}}=\frac{-S_WC_W}{em_Z}C_{h^0\tilde{q}\tilde{q}}~.
\end{eqnarray}
Here $S_W=\sin\theta_W$ and $C_W=\cos\theta_W$  with $\theta_W$ denoting the Weinberg angle. The
coupling constants $C_{{h^0qq}}$ and $C_{{h^0\tilde{q}\tilde{q}}}$ are defined as
\begin{eqnarray}
&&\mathcal{L}\supset C_{h^0qq}(h^0\bar{q}q)+C_{h^0\tilde{q}\tilde{q}}(h^0\tilde{q}\tilde{q}),
\\&&
C_{h^0 dd}=-\frac{1}{\sqrt{2}}Y_dZ^H_{b,1}\;,~~~~~~C_{h^0 uu}=-\frac{1}{\sqrt{2}}Y_uZ^H_{b,2}.
\end{eqnarray}
The couplings of CP-even Higgs with scalar quarks
 $(C_{h^0\tilde{D}_j\tilde{D}_k^*}$ and $C_{h^0\tilde{U}_j\tilde{U}_k^*}$) are deduced as
\begin{eqnarray}
 &&C_{h^0\tilde{D}_j\tilde{D}_k^*}=\frac{1}{12} \sum_{a=1}^{3} \Big\{Z^{D,*}_{j a} Z_{{k a}}^{D}
  \Big(-12 v_d Y_{d,{a}}^2 Z_{{i 1}}^{H}+(v_d Z_{{i 1}}^{H} -v_u Z_{{i 2}}^{H})(3 g_{2}^{2}
   + g_{Y X} g_{X}  + g_{1}^{2} \nonumber \\
 &&+ g_{Y X}^{2})+2  g_{Y X} g_{X}( v_{\eta} Z_{{i 3}}^{H}- v_{\bar{\eta}} Z_{{i 4}}^{H})\Big)+Z^{D,*}_{j 3 + a} Z_{{k 3 + a}}^{D}
 \Big(2 (2 g_{Y X} g_{X}  + 3 g_{X}^{2})( v_{\eta} Z_{{i 3}}^{H}- v_{\bar{\eta}} Z_{{i 4}}^{H}  )\nonumber \\
 &&+(2 g_{1}^{2}+ 2 g_{Y X}^{2}  + 3 g_{X}^{2}  + 5 g_{Y X} g_{X} )(v_d Z_{{i 1}}^{H}
 - v_u Z_{{i 2}}^{H}) -12 v_d Y_{d,{a}}^2 Z_{{i 1}}^{H}\Big)+6 \Big(( Z^{D,*}_{j a}Z_{{k 3 + a}}^{D}
  \nonumber \\
 &&+Z^{D,*}_{j 3 + a}  Z_{{k a}}^{D})[Y_{d,{a}}({\lambda}_{H}v_SZ_{{i 2}}^{H}
  +\sqrt{2}\mu Z_{{i 2}}^{H}+{\lambda}_{H} v_u Z_{{i 5}}^{H})- \sqrt{2}Z_{{i 1}}^{H} T_{d,{a}}]
 \Big)\Big\},
 \\
 &&C_{h^0\tilde{U}_j\tilde{U}_k^*}=\frac{1}{12}\sum_{a=1}^{3} \Big\{Z^{U,*}_{j a} Z_{{k a}}^{U}
 \Big(-12 v_u Y^2_{u,{a}} Z_{{i 2}}^{H}(v_d Z_{{i 1}}^{H} -v_u Z_{{i 2}}^{H})+(  g_{1}^{2} -3 g_{2}^{2}    + g_{Y X} g_{X}\nonumber \\
 && + g_{Y X}^{2})+2  g_{Y X} g_{X} (  v_{\eta} Z_{{i 3}}^{H}- v_{\bar{\eta}} Z_{{i 4}}^{H} )\Big)+Z^{U,*}_{j 3 + a} Z_{{k 3 + a}}^{U}
 \Big(2 (3 g_{X}^{2}  + 4 g_{Y X} g_{X} )(v_{\bar{\eta}} Z_{{i 4}}^{H}-v_{\eta} Z_{{i 3}}^{H} )
 \nonumber\\&&+(3 g_{X}^{2} + 4 g_{1}^{2}  + 4 g_{Y X}^{2}   + 7 g_{Y X} g_{X})(v_u Z_{{i 2}}^{H}-v_d Z_{{i 1}}^{H})
 -12 v_u Y^2_{u,{a}} Z_{{i 2}}^{H} \Big)+6 \Big(( Z^{U,*}_{j a} Z_{{k 3 + a}}^{U}
 \nonumber \\ &&+ Z^{U,*}_{j 3 + a}  Z_{{k a}}^{U})[ Y_{u,{a }}(\sqrt{2}\mu Z_{{i 1}}^{H}
+ {\lambda}_{H} v_d Z_{{i 5}}^{H}
 +{\lambda}_{H}   v_S Z_{{i 1}}^{H})- \sqrt{2}T_{u,{a }}Z_{{i 2}}^{H}] \Big)\Big)\Big\}.
 \end{eqnarray}

For the decay $h^0\rightarrow \gamma\gamma$, the leading order contributions are from the one loop diagrams.
Its decay width is written in the following form
\begin{eqnarray}
&&\Gamma_{U(1)_X}(h^0\rightarrow\gamma\gamma)={G_{F}\alpha^2m_{h^0}^3\over128\sqrt{2}\pi^3}
\Big|\sum\limits_fN_cQ_{f}^2g_{{h^0ff}}A_{1/2}(x_{f})+g_{h^0 H^\pm H^\pm}{m_{ W}^2\over m_{H^\pm}^2}A_0(x_{H^\pm})
\nonumber\\&&
+g_{h^0 WW}A_1(x_{W})
+\sum\limits_{i=1}^2g_{{h^0\chi_i^\pm\chi_i^\pm}}{m_{W}\over m_{{\chi_i}}}A_{1/2}(x_{{\chi_i}})
+\sum\limits_{\tilde f}N_cQ_{f}^2g_{{h^0\tilde{f}\tilde{f}}}{m_{ Z}^2\over m_{{\tilde f}}^2}
A_{0}(x_{{\tilde{f}}})\Big|^2\;.
\label{hpp}
\end{eqnarray}
The function $A_1(x)$ is defined as
\begin{eqnarray}
&&A_1(x)=-\Big[2x^2+3x+3(2x-1)g(x)\Big]/x^2\;.
\end{eqnarray}

The relevant couplings are defined in the following form
\begin{eqnarray}
&&g_{h^0WW}=\frac{S_W}{em_W}C_{h^0WW},~~~~~~~
g_{h^0ZZ}=\frac{C_WS_W}{em_Z}C_{h^0ZZ},\nonumber\\&&g_{h^0\chi_i^\pm\chi_i^\pm}=\frac{-2S_W}{e}C^L_{h^0\chi_i^\pm\chi_i^\pm},~~~
g_{h^0H^\pm H^\pm}=\frac{-v}{2m_W^2}C_{h^0H^\pm H^\pm},\nonumber\\&&
g_{h^0\tilde{L}\tilde{L}}=\frac{-S_WC_W}{em_Z}C_{h^0\tilde{L}\tilde{L}},\\
&&\mathcal{L}\supset C_{h^0WW}g^{\mu\nu}h^0W_\mu W_\nu
+ C_{h^0ZZ}g^{\mu\nu}h^0Z_\mu Z_\nu
+C_{h^0\tilde{L}\tilde{L}}h^0\tilde{L}\tilde{L}\nonumber\\&&
\hspace{0.9cm}+\bar{\chi}_i^\pm(C_{h^0\chi^\pm_i\chi^\pm_j}^LP_L+C_{h^0\chi^\pm_i\chi^\pm_j}^RP_R)\chi_j^\pm h^0+C_{h^0H^\pm H^\pm}h^0H^\pm H^\pm.
\end{eqnarray}

The couplings for $h^0-W-W$ and $h^0-Z-Z$ are
\begin{eqnarray}
 &&C_{h^0WW}=\frac{1}{2} g_{2}^{2} (v_d Z_{{i 1}}^{H}  + v_u Z_{{i 2}}^{H} ),
 \nonumber\\
 &&C_{h^0ZZ}=\frac{1}{2} \Big( [g_1 \cos{\theta'}_W  S_W   + g_2 C_W  \cos{\theta'}_W   - (g_{Y X} + g_{X})\sin{\theta'}_W  ]^{2}\nonumber\\&&\hspace{1.7cm}\times (v_d Z_{{i 1}}^{H} +v_u Z_{{i 2}}^{H}) +4 g_{X}^{2} \sin^{2}{\theta'}_W (v_{\bar{\eta}} Z_{{i 4}}^{H}  + v_{\eta} Z_{{i 3}}^{H} )\Big).
 \end{eqnarray}

 Then CP-even Higgs-slepton-slepton coupling $ C_{h^0\tilde{L}\tilde{L}^*}$ reads as
\begin{eqnarray}
&&C_{h^0\tilde{L}_n\tilde{L}^*_m}=\frac{1}{4}\sum_{a=1}^3\Big\{Z_{m,a}^{E,*}Z_{n,a}^E
\Big((g_2^2-g_{YX}g_X-g_1^2-g_{YX}^2)(v_dZ_{b1}^H-v_uZ_{b2}^H)+g_{YX}g_X(v_{\overline{\eta}}Z_{b4}^H\nonumber\\
&&\hspace{1.6cm}-v_{\eta}Z_{b3}^H)\Big)+Z_{m,3+a}^{E,*}Z_{n,3+a}^E\Big((2g_1^2+2g_{YX}^2+3g_{YX}g_X+g_X^2)(v_dZ_{b1}^H-v_uZ_{b2}^H)\nonumber\\
&&\hspace{1.6cm}+2(g_{YX}g_X+g_X^2)(v_{\eta}Z_{b3}^H-v_{\overline{\eta}}Z_{b4}^H)\Big)+\Big(Z_{m,a}^{E,*}Z_{n,3+a}^E+Z_{m,3+a}^{E,*}Z_{n,a}^E\Big)\nonumber\\
&&\hspace{1.6cm}\times\Big[Y_{e,a}\Big(2(v_S\lambda_H+\sqrt{2}{\mu})Z_{b2}^H+2v_u\lambda_HZ_{b5}^H\Big)-2\sqrt{2}T_{e,a}Z_{b1}^H\Big]\Big\}.
\end{eqnarray}

The CP-even Higgs-$H^\pm$-$H^\pm$ coupling $ C_{h^0H^\pm_m H^{\pm*}_n}$ is
\begin{eqnarray}
&&C_{h^0H^\pm_m H^{\pm*}_n}=\frac{1}{4}\Big\{(-Z_{b2}^HZ_{m2}^+-Z_{b1}^HZ_{m1}^+)\Big([(g_{YX}+g_X)^2+g_1^2+g_2^2]
(v_uZ_{n2}^++v_dZ_{n1}^+)
\nonumber\\
&&+(g_2^2-2\lambda_H^2)(v_dZ_{n1}^+-v_uZ_{n2}^+)\Big)
+(Z_{b2}^HZ_{m1}^++Z_{b1}^HZ_{m2}^+)\Big([(g_{YX}+g_X)^2-2g_2^2+2\lambda_H^2\nonumber\\
&&+g_1^2](v_uZ_{n1}^++v_dZ_{n2}^+)\Big)-2Z_{b4}^H(Z_{m2}^+-Z_{m1}^+)\Big((g_{YX}g_X+g_X^2)v_{\overline{\eta}}(Z_{n2}^++Z_{n1}^+)
+\lambda_cv_{\eta}\lambda_H\nonumber\\&&\times(Z_{n1}^+-Z_{n2}^+)\Big)+Z_{b3}^H(Z_{m2}^++Z_{m1}^+)
(g_{YX}g_X+g_X^2)(v_{\eta}Z_{n1}^+-v_{\eta}Z_{n2}^+)
+\Big[Z_{b4}^H\lambda_cv_{\overline{\eta}}\lambda_H\nonumber\\
&& +Z_{b5}^H\Big(\sqrt{2}T_{{\lambda},H}+2\lambda_H(\kappa v_s+\sqrt{2}M_S
+\sqrt{2}{\mu}+\lambda_Hv_S)\Big)\Big](Z_{m2}^++Z_{m1}^+)(Z_{n2}^++Z_{n1}^+)\Big\}.
\end{eqnarray}

We show the left-handed coupling of $h^0-\chi^\pm_n-\chi^\pm_m$
\begin{eqnarray}
&&  C_{h^0\chi^\pm_n\chi^\pm_m}^L=-\frac{1}{\sqrt{2}}\Big(g_2U_{m1}^*V_{n2}^*Z_{b2}^H
 +U_{m2}^*(g_2V_{n1}^*Z_{b1}^H+\lambda_HV_{n2}^*Z_{b5}^H)\Big).
\end{eqnarray}

The formulae for $h^0\rightarrow ZZ, ~WW$ are expressed as \cite{hzzww1,hzzww2}
\begin{eqnarray}
&&\Gamma(h^0\rightarrow WW)={3e^4m_{{h^0}}\over512\pi^3S_{ W}^4}|g_{h^0WW}|^2
F({m_{ W}\over m_{h^0}}),\;\nonumber\\
&&\Gamma(h^0\rightarrow ZZ)={e^4m_{h^0}\over2048\pi^3S_{W}^4C_{W}^4}|g_{h^0ZZ}|^2
\Big(7-{40\over3}S_{W}^2+{160\over9}S_{W}^4\Big)F({m_{Z}\over m_{h^0}}).
\end{eqnarray}
The concrete form of $F(x)$ is
\begin{eqnarray}
&&F(x)=-(1-x^2)\Big({47\over2}x^2-{13\over2}+{1\over x^2}\Big)-3(1-6x^2+4x^4)\ln x
\nonumber\\
&&\hspace{1.5cm}
+{3(1-8x^2+20x^4)\over\sqrt{4x^2-1}}\cos^{-1}\Big({3x^2-1\over2x^3}\Big)\;.\nonumber\\
\label{form-factor1}
\end{eqnarray}

\begin{figure}[h]
\setlength{\unitlength}{1mm}
\centering
\includegraphics[width=7in]{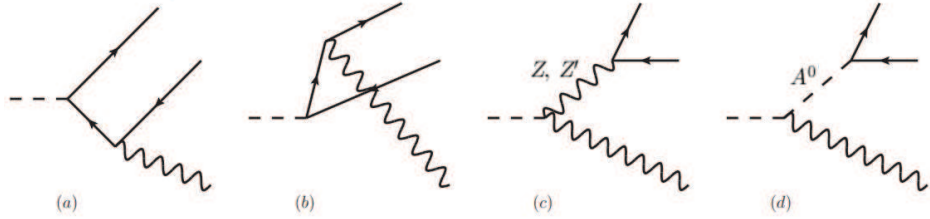}
\vspace{-20.3cm}
\caption[]{The Feynman diagrams for $h^0\to\bar{l}lZ.$}\label{hllzfmtu}
\end{figure}

 We also study the processes of $h^0\to\bar{l}lZ$ and $h^0\to\bar{\nu}_l\nu_lZ~({\rm with}~l=e,~\mu,~\tau)$.
The latter is simpler than the former, and can be obtained by taking the limit $m_l\to 0$ from the former, where $m_l$ denotes the mass of lepton.
For the processes of $h^0(p_1)\to l(p_2)+\bar{l}(p_3)+Z(p_4)$, the diagrams are shown in the Fig.\ref{hllzfmtu}.
The external particles all satisfy the on shell condition: $p_1^2=m_{h^0}^2,~p_2^2=p_3^2=m_l^2$ and
$p_4^2=m_Z^2$. We use the Mandelstam invariants: $s=(p_2+p_3)^2,~t=(p_3+p_4)^2,~u=(p_2+p_4)^2$ and
$s+t+u=m_{h^0}^2+2m_l^2+m_Z^2$. In our calculation, all lepton masses and the couplings of $\bar{l}-H^0(G^0)-l$ are kept.

From the diagrams in the Fig.\ref{hllzfmtu}, we can obtain the decay width through the following formula\cite{QCF}
\begin{eqnarray}
\Gamma(h^0\rightarrow l\bar{l}Z)=\frac{1}{256\pi^3m^3_{h^0}}
\int_{4m_l^2}^{(m_{h^0}-m_Z)^2}ds\int_{t^-}^{t^+}\sum|\mathcal{M}|^2.
\end{eqnarray}
Here, $\mathcal{M}$ is the Feynman amplitude for the Fig.\ref{hllzfmtu}. The definitions of $t^\pm$ are
\begin{eqnarray}
t^\pm=\frac{1}{2}\Big[m_{h^0}^2+2m_l^2+m_Z^2-s\pm\Big(1-\frac{4m_l^2}{s}\Big)^{1/2}\Big((m_{h^0}^2+m_Z^2-s)^2-4m^2_{h^0}m_Z^2\Big)^{1/2}\Big].
\end{eqnarray}

\section{numerical results}
   To study the lightest CP-even Higgs $h^0$ mass ($m_{h^0}\simeq$125 GeV)
  and $h^0$ decays $h^0\rightarrow VV ~({\rm with}~ V=\gamma,~W,~Z)$,
  $h^0\rightarrow l\bar{l}Z$ and $h^0\rightarrow \nu_l\bar{\nu}_lZ~({\rm with}~l=e,~\mu,~\tau)$ in the $U(1)_X$SSM,
  we consider the mass constraint for the $Z^\prime$ boson ($M_{Z^{\prime}}> 5.1$ TeV)\cite{Zp5d1} from LHC experiments.
The constraints $M_{Z^\prime}/g_X\geq 6 ~{\rm TeV}$\cite{UPbmzgx} and $\tan \beta_\eta< 1.5$\cite{TBnew} are also taken into account.
The parameters are used to make the scalar lepton masses larger than 700 GeV,
and chargino masses larger than 1100 GeV\cite{SUSYmass}.

  Some parameters are adopted here with $i=1,2,3$
\begin{eqnarray}
&&\kappa= Y_{Xii}=1, ~v_S=3.6~{\rm TeV},  ~\mu=M_1=T_{d ii}=1~{\rm TeV},~M_S=2.7~{\rm TeV},
\nonumber\\&&\tan{\beta_\eta}=0.8, ~
B_\mu=B_S= m_{\tilde{L}_{ii}}^2=1~{\rm TeV^2},~ l_W=4~{\rm TeV^2},~l_S=-300~{\rm TeV^3},\nonumber\\&&
T_\kappa=1.6~{\rm TeV},~M_2=1.2~{\rm TeV},~M_{BL}=0.3~{\rm TeV},~
T_{Xii}=T_{eii}=0.5~{\rm TeV},
\nonumber\\&&
T_{\nu ii}=0.8~{\rm TeV},~m_{\tilde{E}_{ii}}^2= m_{\tilde{D}_{ii}}^2=5~{\rm TeV^2},~m_{\tilde{\nu}_{ii}}^2=0.5~{\rm TeV^2},~
\xi=17~{\rm TeV}.\label{canshu}
\end{eqnarray}

To simplify the numerical discussion, we use the following relations
\begin{eqnarray}
&&T_{uii}=T_u, ~m_{\tilde{U}ii}^2=M_{U}^2, ~m_{\tilde{Q}ii}^2=M_{Q}^2, ~(i=1,2,3).
\end{eqnarray}
The parameters $T_u, m_{\tilde{Q}}$ and $m_{\tilde{U}}$ all emerge in the soft breaking terms and are included in $\mathcal{L}_{soft}^{MSSM}$.
$T_u$ is the coupling constant for trilinear scalar coupling $T_uH_u\tilde{Q}\tilde{U}$.
For the scalar quark fields $\tilde{Q}$ and $\tilde{U}$, $m_{\tilde{Q}}$ and $m_{\tilde{U}}$ are respectively mass terms corresponding to $m_{\tilde{Q}^2}\tilde{Q}^*\tilde{Q}$ and $m_{\tilde{U}^2}\tilde{U}^*\tilde{U}$.

To explore the parameter space better, we randomly scan the parameters as the following
\begin{eqnarray}
&&5\leq\tan\beta\leq50,~~~ 0.3\leq g_X\leq0.8, ~~~0.01\leq g_{YX}\leq0.5,~~~
-1\leq\lambda_C\leq1,
 \nonumber\\&&-1\leq\lambda_H\leq1,~~ ~-2~{\rm TeV}\leq T_{\lambda_C}\leq2~{\rm TeV},~~~
-2~{\rm TeV}\leq T_{\lambda_H}\leq2~{\rm TeV},
\nonumber\\&&-5~{\rm TeV}\leq T_{u}\leq5~{\rm TeV}, ~4~{\rm TeV}^2\leq M_{Q}^2\leq8~{\rm TeV}^2, ~
4~{\rm TeV}^2\leq M_{U}^2\leq8~{\rm TeV}^2.
\end{eqnarray}

\begin{figure}[h]
\setlength{\unitlength}{1mm}
\centering
\includegraphics[width=2.9in]{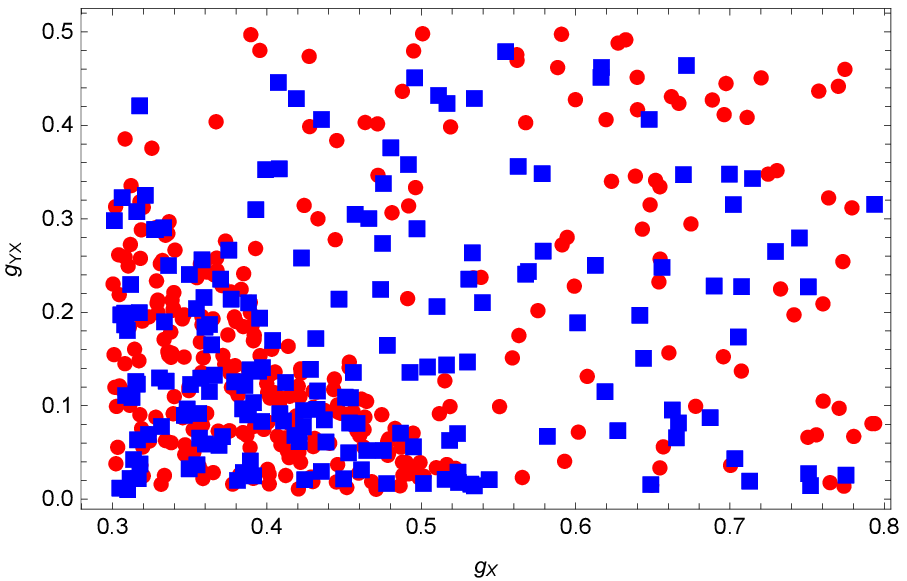}~~~\includegraphics[width=2.9in]{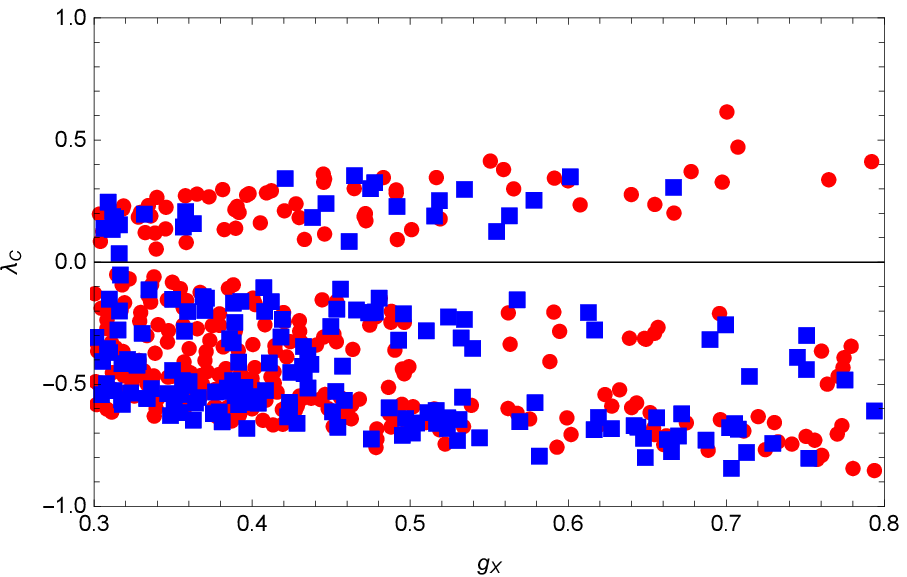}
\caption[]{For the left diagram, the lightest CP-even Higgs mass($m_{h^0}$) in the plane of $g_X$ versus $g_{YX}$;
for the right diagram, $m_{h^0}$ in the plane of $g_X$ versus $\lambda_{C}$.}\label{Mhiggstu1}
\end{figure}

\begin{figure}[h]
\setlength{\unitlength}{1mm}
\centering
\includegraphics[width=2.9in]{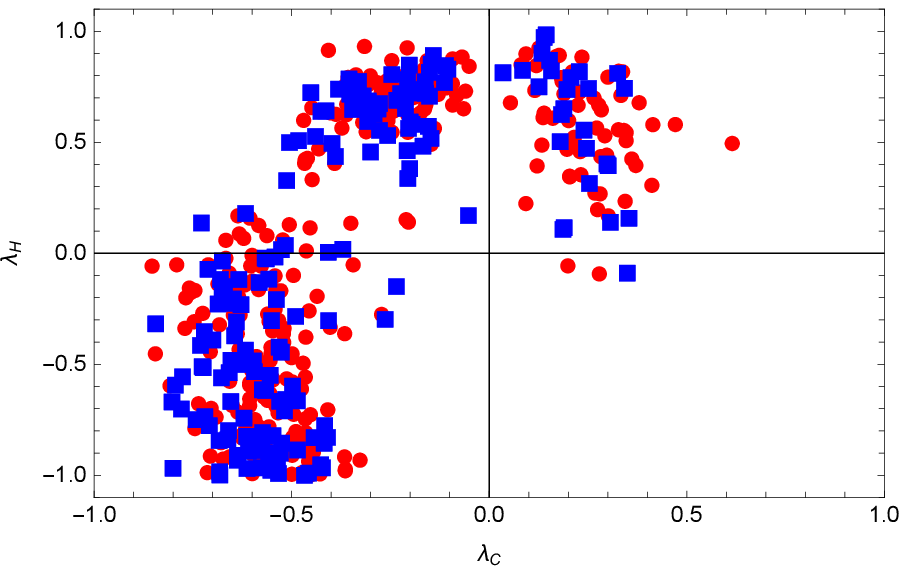}~~~\includegraphics[width=2.9in]{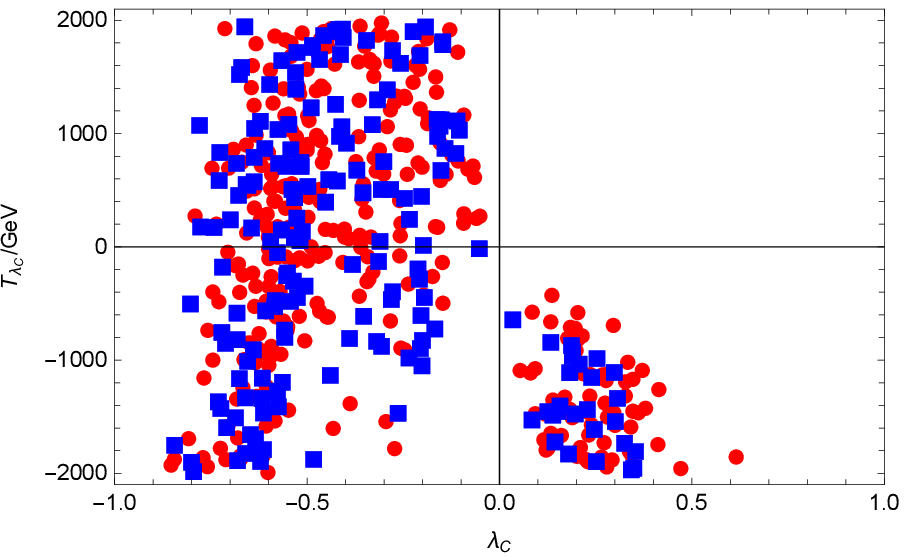}
\caption[]{For the left diagram, $m_{h^0}$ in the plane of $\lambda_C$ versus $\lambda_H$;
for the right diagram, $m_{h^0}$ in the plane of $\lambda_{C}$ versus $T_{\lambda_{C}}$.}\label{Mhiggstu2}
\end{figure}

In the Fig.\ref{Mhiggstu1} and Fig.\ref{Mhiggstu2},
 \textcolor{blue}{$\blacksquare$} denote the lightest CP-even Higgs mass in the region $123\;{\rm GeV}\leq m_{h^0}\leq 127\;{\rm GeV}$.
\textcolor{red}{$\bullet$} represent the regions $120\;{\rm GeV}\leq m_{h^0}< 123\;{\rm GeV}$ and
$127\;{\rm GeV}< m_{h^0}\leq 130\;{\rm GeV}$. In the left diagram of Fig.\ref{Mhiggstu1},
the numerical results of $m_{h^0}$ are plotted in the plane of $g_X$ versus $g_{YX}$.
$g_X$ is the gauge coupling constant of the $U(1)_X$ group.
$g_{YX}$ is the mixing gauge coupling constant of  $U(1)_Y$ group and $U(1)_X$ group.
So they should give considerable effects to the results. Most of \textcolor{blue}{$\blacksquare$} and \textcolor{red}{$\bullet$}
are concentrated at bottom left part, which form a
triangle with two slides as $0.3\leq g_X\leq0.55$ and $0< g_{YX}\leq0.35$.
The points are sparse in the other region. For the right diagram, we show the results of $m_{h^0}$ in the plane of $g_X$ versus $\lambda_{C}$.
The points in the region $\lambda_C<0$ are more than those in the region $\lambda_C>0$.
The concentrated area of \textcolor{blue}{$\blacksquare$} and \textcolor{red}{$\bullet$} is like a rectangle with $-0.1\geq\lambda_C\geq-0.6$ and $0.3\leq g_X\leq0.45$.

In the left diagram of Fig.\ref{Mhiggstu2}, $m_{h^0}$ is shown in the plane of $\lambda_C$ versus $\lambda_H$.
Many points are in the first, second and third quadrants. In the fourth quadrant, so few points appear.
It implies that the second and third quadrants are better for the \textcolor{blue}{$\blacksquare$}.
The numerical results of $m_{h^0}$ versus $\lambda_{C}$ and $T_{\lambda_{C}}$ are plotted
in the right diagram of Fig.\ref{Mhiggstu2}. Obviously, the first quadrant is blank.
That is to say, as $\lambda_{C}>0$ and $T_{\lambda_{C}}>0$, there is not any suitable result.
A lot of \textcolor{blue}{$\blacksquare$} and \textcolor{red}{$\bullet$} emerge in the second and third quadrants.
In the fourth quadrant, the points concentrate at the lower left corner.
From Figs.\ref{Mhiggstu1} and \ref{Mhiggstu2}, we can find that $\lambda_{C}$,
$\lambda_{H}$ and $T_{\lambda_{C}}$ are sensitive parameters for $m_{h^0}$.

\begin{figure}[h]
\setlength{\unitlength}{1mm}
\centering
\includegraphics[width=2.9in]{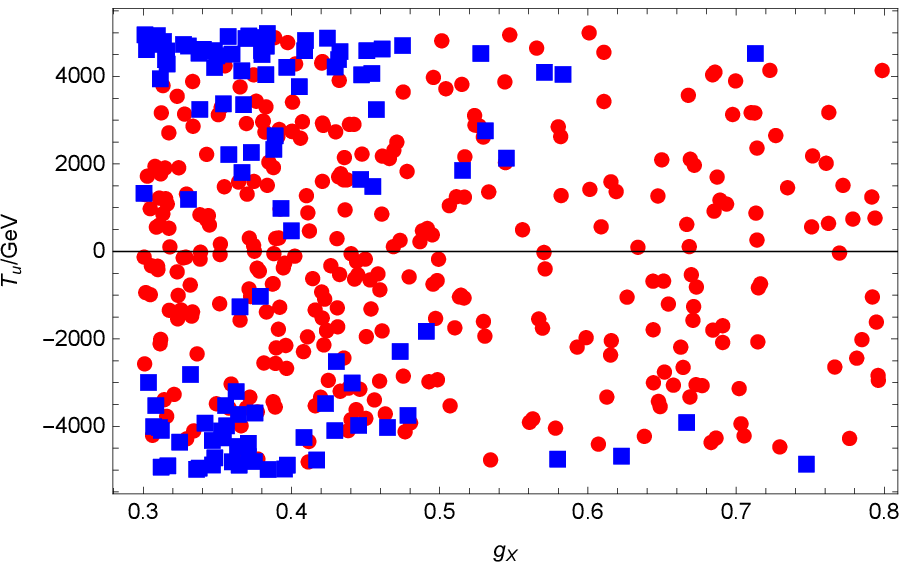}~~~\includegraphics[width=2.9in]{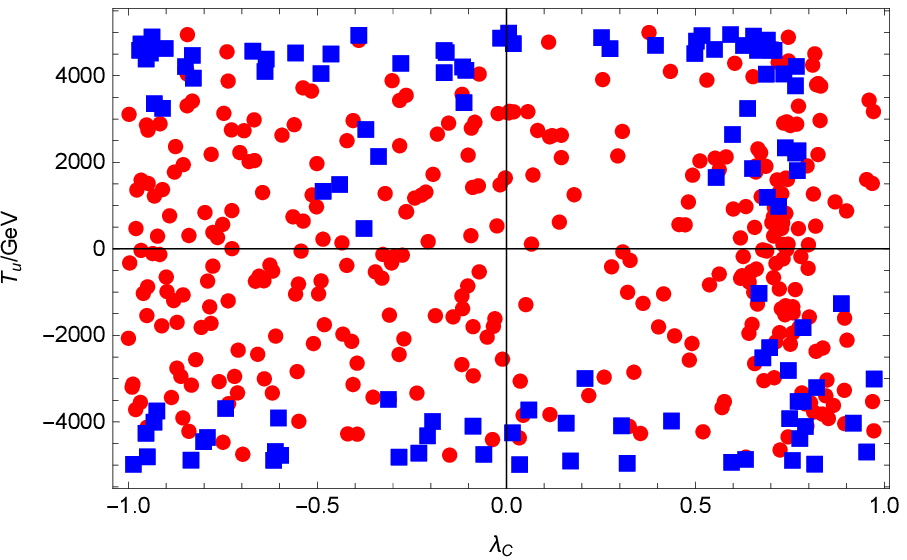}
\caption[]{For the left diagram, $R_{\gamma\gamma}$ in the plane of $g_X$ versus $T_u$;
for the right diagram, $R_{\gamma\gamma}$ in the plane of $\lambda_{C}$ versus $T_u$.}\label{2gammatu1}
\end{figure}

\begin{figure}[h]
\setlength{\unitlength}{1mm}
\centering
\includegraphics[width=2.9in]{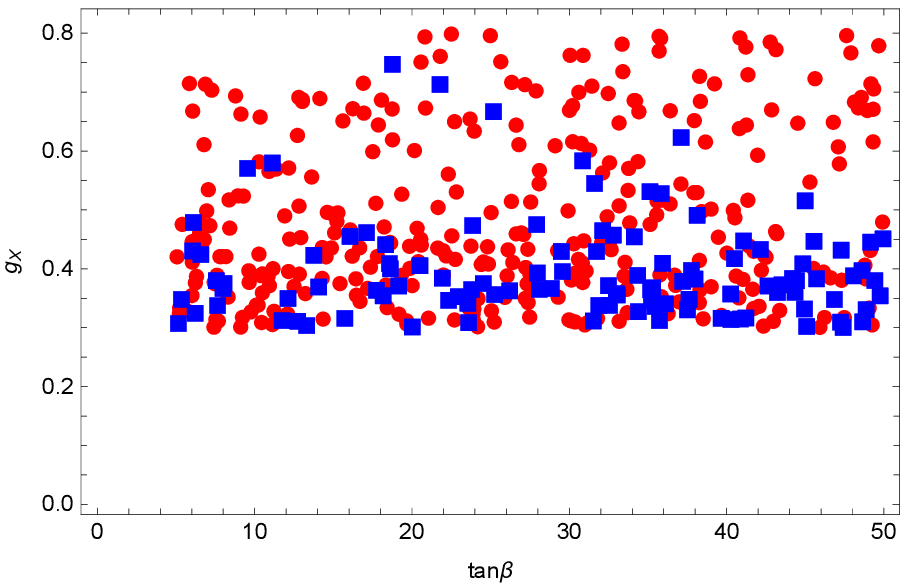}~~~\includegraphics[width=2.9in]{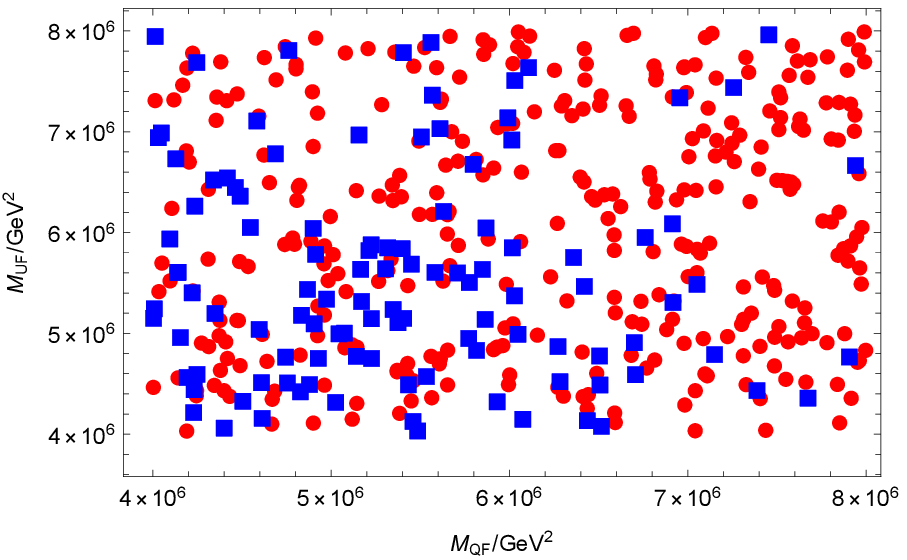}
\caption[]{For the left diagram, $R_{\gamma\gamma}$ in the plane of $\tan\beta$ versus $g_X$;
for the right diagram, $R_{\gamma\gamma}$ in the plane of $M_{QF}$ versus $M_{UF}$.}\label{2gammatu2}
\end{figure}

The ratio $R_{\gamma\gamma}$ is also researched,
and the corresponding results are shown in the
Fig.\ref{2gammatu1} and Fig.\ref{2gammatu2}, where the notations are
\textcolor{red}{$\bullet$}$\rightarrow 0.95\leq R_{\gamma\gamma}\leq1.03$ and \textcolor{blue}{$\blacksquare$}
$\rightarrow 1.03< R_{\gamma\gamma}\leq1.17$.
These points \textcolor{blue}{$\blacksquare$} and
 \textcolor{red}{$\bullet$} satisfy the constraint from $m_{h^0}$ with $124~{\rm GeV} \leq m_{h^0} \leq 126 ~{\rm GeV}$.
 The left diagram of Fig.\ref{2gammatu1} embodies $R_{\gamma\gamma}$ in the plane of $g_X$ versus $T_u$.
 Many points appear in the region $0.3\leq g_X \leq 0.5$,
 where \textcolor{red}{$\bullet$} concentrate in the $T_u$ range as $-3000~{\rm GeV}\leq T_u\leq3000~{\rm GeV}$
 and \textcolor{blue}{$\blacksquare$} are distributed on the upper and lower sides of \textcolor{red}{$\bullet$}.
The effects from $\lambda_{C}$ and $T_u$ to $R_{\gamma\gamma}$ are studied in the right diagram of Fig.\ref{2gammatu1}, where
\textcolor{blue}{$\blacksquare$} and \textcolor{red}{$\bullet$} are distributed almost the entire area of the graph.
\textcolor{blue}{$\blacksquare$} mainly appear in the areas above, below, and to the right.

Both $\tan\beta$ and $g_X$ influence $R_{\gamma\gamma}$, which is shown by the left diagram of Fig.\ref{2gammatu2}.
It implies that $\tan\beta$ is an insensitive parameter and the effect from $\tan\beta$ is mild.
The \textcolor{blue}{$\blacksquare$} and \textcolor{red}{$\bullet$} are concentrated in the area $0.3\leq g_X\leq0.5$.
In the right diagram of Fig.\ref{2gammatu2}, the numerical results of $R_{\gamma\gamma}$ are plotted in the plane of $M_{QF}$ versus $M_{UF}$.
$M_{QF}$ and $M_{UF}$ affect scalar quark mass, which affect $m_{h^0}$ and $R_{\gamma\gamma}$.
The \textcolor{red}{$\bullet$} are distributed throughout the region of the figure.
In the upper right corner, there is almost no \textcolor{blue}{$\blacksquare$}.
Smaller $M_{QF}$ and $M_{UF}$ produce relative light scalar quarks,
which can improve the scalar quark contributions to $R_{\gamma\gamma}$.
Therefore, many \textcolor{blue}{$\blacksquare$} emerge at the lower left corner.

\begin{figure}[h]
\setlength{\unitlength}{1mm}
\centering
\includegraphics[width=2.9in]{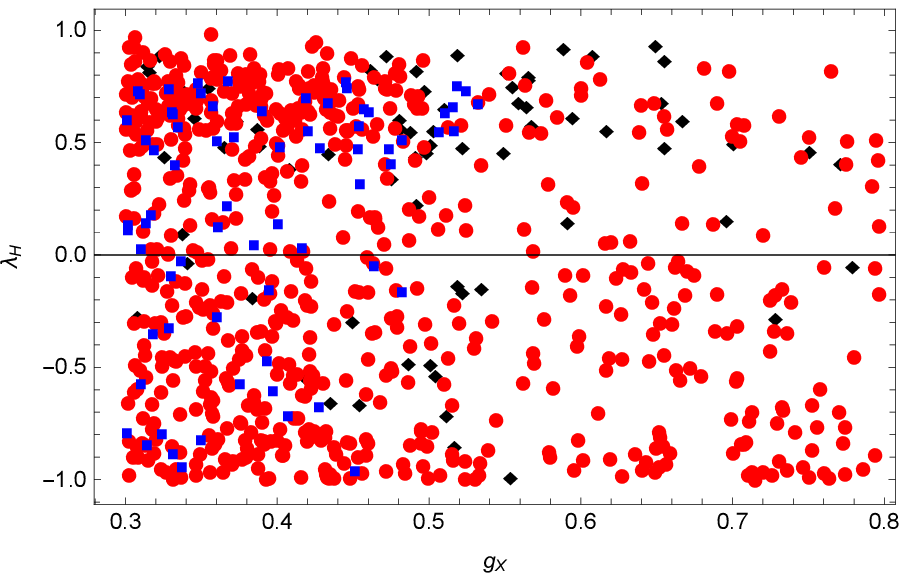}~~~\includegraphics[width=2.9in]{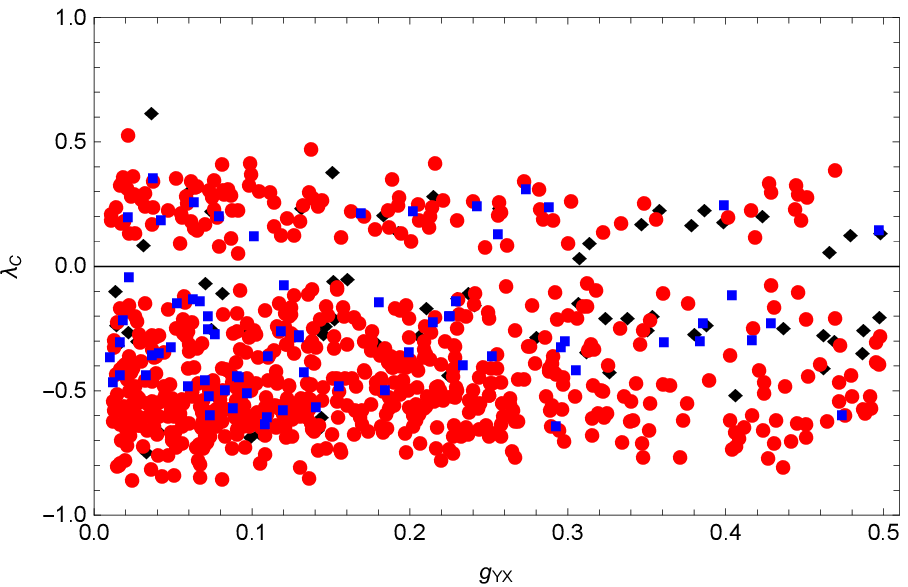}
\caption[]{For the left diagram, $R_{VV}~({\rm with} ~V=W,~Z)$ in the plane of $g_X$ versus $\lambda_H$;
for the right diagram, $R_{VV}$ in the plane of $g_{YX}$ versus $\lambda_{C}$.}\label{2VVtu}
\end{figure}

Similarly, we calculate the decays $h^0\rightarrow WW$ and $h^0\rightarrow ZZ$. From the numerical results,
we find $R_{WW}$ is very close to $R_{ZZ}$. Therefore, we use $R_{VV}~({\rm with}~V=W,~Z)$ to denote the both ratios.
In the Fig.\ref{2VVtu}, the results of $R_{VV}$ are plotted with $\blacklozenge$, \textcolor{red}{$\bullet$} and \textcolor{blue}{$\blacksquare$}.
Here, $\blacklozenge$  represent $0.9\leq R_{VV}<1$, \textcolor{red}{$\bullet$} represent $1\leq R_{VV}\leq1.03$, and \textcolor{blue}{$\blacksquare$} denote $1.03\leq R_{VV}\leq1.20$.
The left diagram of the Fig.\ref{2VVtu} shows the relation between $R_{VV}, g_X$ and $\lambda_H$.
Most \textcolor{red}{$\bullet$} and all \textcolor{blue}{$\blacksquare$} are concentrated in the
region $0.3\leq g_X\leq0.55$, which implies that large $g_X$ is not favourable.
The number of \textcolor{blue}{$\blacksquare$} is smaller than that of \textcolor{red}{$\bullet$}.
For the right diagram, the points concentrate in the region $-1<\lambda_C<0$ and $0<g_{YX}<0.28$.
As $\lambda_C>0.5$, there is not suitable point. For \textcolor{blue}{$\blacksquare$},
the region $-0.7<\lambda_C<0$ and $0<g_{YX}<0.15$ is advantageous.
In the both diagrams, $\blacklozenge$ are not a lot, and \textcolor{red}{$\bullet$} are dominant.

Here, we study the Higgs boson decays $h^0\rightarrow l\bar{l}Z$ and $h\rightarrow \nu_l\bar{\nu}_l Z~({\rm with}~l=e,~\mu,~\tau)$
with the parameters in Eq.(\ref{canshu}). The other used parameters are
\begin{eqnarray}
&&\tan\beta=10,~~~ g_X=0.37, ~~~ g_{YX}=0.1,~~~
\lambda_C=-0.1,
 \nonumber\\&&\lambda_H=0.6, ~~~T_{\lambda_C}=-0.1~{\rm TeV},~~~
T_{\lambda_H}=0.3~{\rm TeV}.
\end{eqnarray}
After calculation, we obtain the corresponding numerical results in the $U(1)_XSSM$
\begin{eqnarray}
&&Br(h^0\rightarrow l\bar{l}Z)=8.1\times10^{-4},~~~l=e,~\mu,\nonumber\\&&
Br(h^0\rightarrow \tau\bar {\tau}Z)=8.5\times10^{-4},\nonumber\\&&
Br(h^0\rightarrow \nu_l\bar{\nu}_l Z)=1.6\times10^{-3},~~~l=e,~\mu,~\tau. \label{Rllz}
\end{eqnarray}
For the above decays, the numerical results in Eq.(\ref{Rllz}) are stable.
 The authors study these decays in the SM, and give the numerical results as\cite{QCF}
 \begin{eqnarray}
&&Br(h^0\rightarrow l\bar{l}Z)=7.5\times10^{-4},~~~l=e,~\mu,\nonumber\\&&
Br(h^0\rightarrow \tau\bar {\tau}Z)=7.3\times10^{-4},\nonumber\\&&
Br(h^0\rightarrow \nu_l\bar{\nu}_l Z)=1.5\times10^{-3},~~~l=e,~\mu,~\tau. \label{QCFRllz}
\end{eqnarray}
Comparing with the branching ratios in Eq.(\ref{QCFRllz}), our numerical results are of the same order
and a little bigger than their results. This characteristic should be caused by the new physics contributions.
From our numerical research,  we find that the processes concerned are attainable in the LHC experiments
and may be detected in the near future.

\section{discussion and conclusion}
 Introducing three Higgs singlets and right-handed neutrinos to the $U(1)_X$ extension of MSSM,
we obtain the $U(1)_X$SSM. In the $U(1)_X$SSM, the neutral CP-even parts of two Higgs doublets($H_d$ and $H_u$)
and tree Higgs singlets ($\eta,~\bar{\eta}$ and $S$) mix together, which constitute a $5\times5$ mass squared matrix of CP-even Higgs.
The lightest eigenvalue corresponds to $m_{h^0}$, but at tree level it can not reach 125 GeV.
The loop corrections should be taken into account.
In this work, we use the Higgs effective potential with one loop corrections to study the Higgs mass $m_{h^0}$.  The constraint
from $m_{h^0}$ near 125 GeV confines the parameter space obviously. The loop corrections from scalar top quark are dominant among the
SUSY loop corrections.

The Higgs boson decays $h^0\rightarrow \gamma\gamma$ and $h^0\rightarrow VV~({\rm with}~V=W,~Z)$ are calculated.
From the numerical results for $R_{\gamma\gamma}$ and $R_{VV}$ ( with $V=W,Z$ ),
 we find that the contributions of the $U(1)_X$SSM are visible and can make these ratios
 larger than 1 in the reasonable parameter spaces.
 Our numerical results are closer to the experimental data than the corresponding predictions of the SM.
 For the researched Higgs boson decays $h^0\rightarrow l\bar{l}Z~({\rm with}~l=e,~\mu,~\tau)$, the branching ratios are in the region $(8\sim9)\times 10^{-4}$.
The branching ratios $R(h^0\rightarrow \nu_l\bar{\nu}_l Z)~({\rm with}~l=e,~\mu,~\tau)$ are also calculated, whose numerical results are around
$1.6\times10^{-3}$.  For the Higgs boson decays $h^0\rightarrow l\bar{l}Z$ and
 $h^0\rightarrow \nu_l\bar{\nu}_l Z~({\rm with}~l=e,~\mu,~\tau)$, our numerical
 results of their branching ratios are a little bigger than the results in Ref.\cite{QCF}.
In the order analysis, these branching ratios are not small, and they are in the detectable range of LHC\cite{QCF,exp1,exp2,exp3,exp4}.
 Studying the new physics contributions to the Higgs boson rare decays is useful for testing
 the Higgs boson property and searching for new physics beyond the SM.
 We hope that these decays will be detected in the near future, and they benefit the study of Higgs boson.

\begin{acknowledgments}
This work is supported by National Natural Science Foundation of China (NNSFC)
(No. 12075074), Natural Science Foundation of Hebei Province
(A2020201002, A202201022, A2022201017), Natural Science Foundation of Hebei Education Department(QN2022173).
\end{acknowledgments}

\end{document}